\documentclass[amsmath,amssymb,aps,prd,reprint,notitlepage,twocolumn,showpacs]{revtex4-1}
\usepackage{graphicx}% Include figure files
\usepackage{hyperref}% add hypertext capabilities
\usepackage[titletoc,title]{appendix}
\usepackage{siunitx}

\begin{document}
\title{Dark Matter near gravitating bodies}
\author{H. B. Tran Tan}
\author{V. V. Flambaum}
\author{J. C. Berengut}
\affiliation{School of Physics, University of New South Wales, Sydney, NSW 2052, Australia}
\date{\today}

\begin{abstract}
\begin{center}
\textbf{\textit{Abstract}}\\
\end{center}
\textit{In this paper, we show that in the vicinity of certain astronomical bodies, e.g., a Neutron Star, a Black Hole, there exist significant enhancements of Dark Matter's density and current, due to its interaction with the gravitational field of the bodies. This enhancement implies that the effects of Dark Matter - Normal Matter interactions are enhanced and hence might be observable.}
\end{abstract}

\maketitle

\section{Introduction}
One of the gravest difficulties encountered when searching for Dark Matter (DM) is the fact that its non-gravitational interactions with Standard Model Matter (SMM) are very weak. For example, the current limit on the axion-photon coupling constant is $g_{a\gamma} \lesssim \SI{e-11}{\per\giga\electronvolt}$ (axion is a prominent DM candidate; for a review of the QCD axion, see \cite{Peccei2008}; for a review of the axion's role in cosmology and astrophysics, see \cite{Sikivie2008,Raffelt2008,MARSH20161}), the current limit on the axion-electron coupling constant is $g_{ae} \lesssim \SI{e-13}{\per\giga\electronvolt}$. In fact, the DM-SMM interactions are so small that all DM searching experiments have yet to yield positive results. Therefore, it is of great interest to find situations in which these interactions are enhanced. %Because of this, proposals for DM detection have been concentrating on finding ways to enhance the observable effects of these interactions. One such way is to use laser and maser spectroscopy to search for DM-induced variation of the fine structure constant $\alpha=\frac{e^2}{4\pi}$ \cite{stadnik2015enhanced}: it was shown that the change of phase difference $\varphi$ (an observable) between lights traveling along the two arms of the spectroscopy is related to the change of $\alpha$ by $\delta \varphi = N \delta \alpha$ where $N$ is a number of the order of $10^{14} \sim 10^{16}$. 

Since the DM-SMM interactions are proportional to the density of interacting particles, one expects to observe enhancements of these interactions where DM is abundant. Because gravitational interaction is attractive and is, apparently, the strongest kind of interaction between DM and SMM, the DM abundance should be high near some heavy and relatively small astronomical object, where the gravitational field is strong. Naturally, White Dwarfs, Neutron Stars and Black Holes are good candidates. On the other hand, since Earth is relatively lightweight for its size, one expects the DM density near Earth to be the same as that in the void between astronomical bodies.

Similarly, since some DM-SMM interactions depend on the DM particle current, one is also interested in the enhancement of this quantity. This enhancement amplifies, for example, the absorption of DM particles by atoms and molecules (recall that the absorption rate is the product of the absorption cross-section and the flux i.e., current strength of the incoming DM particles) and the effects on SMM particles by the pseudo-magnetic field created by a DM current.

In this paper, we calculate and compare the DM particle density and current near Earth, Sun, a typical Neutron Star, and a typical White Dwarf. %and the stars S1 when it comes closest to the Supermassive Black Hole Sagittarius-A*.

\section{Problem Set-up}
The behavior of DM particles in the gravitational potential due to some mass distribution (body) can be studied in the framework of classical mechanics. This problem was considered in \cite{SikivieWick2002}. The DM particles considered therein are either WIMPs with mass in the range $10$ to $\SI{100}{\giga\electronvolt}$ or axions with mass in the range $10^{-5}$ to $\SI{e-6}{\electronvolt}$ and the bodies considered were Sun and Earth. Since the typical velocity of the DM particles relative to these bodies is a few hundred $\SI[per-mode=symbol]{}{\kilo\meter\per\second}$ (the Sun's velocity is $\SI[per-mode=symbol]{220}{\kilo\meter\per\second}$, the DM ensemble's root-mean-square velocity is $\SI[per-mode=symbol]{270}{\kilo\meter\per\second}$), the de Broglie wavelengths of these particles were $10^{-17}\sim\SI{e-16}{\kilo\meter}$ for WIMPs and $1\sim\SI{10}{\kilo\meter}$ for axions, much less than the radii of Sun or Earth. Thus, the classical mechanics consideration was proper.  The analysis in \cite{SikivieWick2002} can be extended to the cases of Neutron Stars and White Dwarfs. We will compare the results of these extensions with our results in the following sections.

If one chooses to consider DM particle whose mass $\mu$ is so small such that the corresponding de Broglie wavelength $\lambda_B$ is greater than the size of the body in question, the behavior of the DM particles near the body is quantum mechanical in nature. 
In Table \ref{de Broglie mass}, we present the upper bound of $\mu$ such that this is the case. The sizes of the bodies question are taken to be their radii $R$. The typical velocity of a DM particle relative to the bodies in question is taken to be $\SI[per-mode=symbol] {500}{\kilo\meter\per\second}$ (we will not be too concerned about the actual value; the figure assumed here is in the correct order of magnitude). The upper bound on $\mu$ is defined by the condition $\lambda_B \geq 10R$.
\begin{table}[htb]
\small
\centering
\begin{tabular}{|>{\centering\arraybackslash}p{1.2cm}|c|c|>{\centering\arraybackslash}p{1.8cm}|>{\centering\arraybackslash}p{1.8cm}| }
\hline
\textbf{Object} & $M$ $\left(\SI{}{\kilogram}\right)$ & $R$ $\left(\SI{}{\kilo\meter}\right)$ & \textbf{Upper bound for} $\mu$ $\left(\SI{}{\electronvolt}\right)$ & \textbf{Upper bound for} $\kappa = GM\mu$ \\
\hline
Earth & $\SI{6e24}{}$ & 6371 & $\SI{e-11}{}$ & $\SI{3e-7}{}$\\
\hline
Sun & $\SI{2e30}{}$ & 695842  & $\SI{e-13}{}$ & $\SI{8e-4}{}$\\
\hline
Neutron star & $\SI{4e30}{}$ & 12 & $\SI{6e-9}{}$ & 90 \\
\hline
White Dwarf & $\SI{2e30}{}$ & 6371 & $\SI{e-11}{}$ & 0.09 \\
\hline
%Sgr-A* & $\SI{8.11e36}{}$ & $6.7 \times 10^9$ & $\SI{2.04e-17}{}$ & 0.64\\
%\hline
\end{tabular}
\caption{Maximal value of the DM mass $\mu$ such that the DM particles behave quantum mechanically in the potential of some astronomical objects. The corresponding bounds on the coupling strength $\kappa = GM\mu$ ($M$ is the mass of the body) are also presented. In this paper, only DM particles with masses less than the bounds given here are considered.} \label{de Broglie mass}
\end{table}

For definiteness, in the calculations below, we assume these bounds for the masses of the DM particles in question. However, the final results will not depend on these masses, as to be expected in classical mechanics (gravity affects all particles in the same way).

Another feature of the classical calculation \cite{SikivieWick2002} is the consideration of the particles whose orbits pass through the mass distribution at some point. This leads to the replacement of a pure Coulomb potential by one with a non-Coulomb behavior inside the body. This correction to the potential changes the results significantly compare to the case of a point mass.

In our consideration, the difference between a mass distribution of finite size and a point mass is not crucial. Recall that the partial wave of angular momentum $l$ is scattered significantly by a potential of range $R$ only if $l \leq kR$ where $k$ is the momentum of the scattered particle. Thus, if it happens that $kR <1$, the potential has no substantial effect on the particle. As mentioned above, the condition for the applicability of quantum mechanics to the current problem is $\lambda_B > R$. This condition implies $kR <1$. Therefore, we are justified in assuming that the mass distributions are point-like and the gravitational potentials are purely Coulomb-like.

Also, in this paper, we neglect all non-gravitational couplings, since we assume that they are very weak compared to gravity, and the gravitational interactions among DM particles, since the mass of these particles are very small compared to that of the bodies in question. 

With these simplifications, the situation reduces to the followings: each DM particle scatters off the Coulomb-like gravitational potential of the body independently from one another. The (scalar) wavefunction $\psi$ of a non-relativistic quantum mechanical particle scattering off a Coulomb potential is well known \cite{landau1958course}. From this wavefunction, we obtain the \textit{single} particle density, which reads
%\begin{widetext}
\begin{equation}\label{one particle density}
\begin{aligned}
	\rho \left( \mathbf{r},\mathbf{v} \right) &= \left| \psi \right|^2=			\frac{2\pi \kappa }{v\left( 1-{{e}^{-\frac{2\pi \kappa }{v}}} \right)}\\
	& \times {{\left| K\left( \frac{i\kappa }{v},1,i\mu \left( vr-\mathbf{v}\cdot \mathbf{r} \right) \right) \right|}^{2}} \,,
\end{aligned} 
\end{equation}
%\end{widetext}
and the \textit{single} particle current, which reads
%\begin{widetext}
\begin{equation}\label{one particle current}
\begin{aligned}
  \mathbf{j}&=\frac{\operatorname{Im}\left( {{\psi }^{*}}\nabla \psi  \right)}{\mu }\\
  &=\frac{2\pi \kappa \mathbf{v}}{v\left( 1-{{e}^{-\frac{2\pi \kappa }{v}}} \right)}\left({{\left| K\left( \frac{i\kappa }{v},1,i\mu \left( vr-\mathbf{v}\cdot \mathbf{r} \right) \right) \right|}^{2}} \right.\\
  &+\left.B\left(\frac{i\kappa}{\mu},i\mu \left( vr-\mathbf{v}\cdot \mathbf{r} \right)\right)\right) \\
  &-\frac{2\pi {{\kappa }}{{\mathbf{e}}_{r}}}{\left( 1-{{e}^{-\frac{2\pi \kappa }{v}}} \right)}B\left(\frac{i\kappa}{\mu},i\mu \left( vr-\mathbf{v}\cdot \mathbf{r} \right)\right) \,,
\end{aligned}
\end{equation}
%\end{widetext}
where $\kappa =GM\mu $ ($M$ is the body's mass), $\mathbf{v}$ is the DM particle's velocity at infinity, $\mathbf{r}$ is the vector from the scattering center (the body) to the position at which the density and current are measured, $\mathbf{e}_r$ is the unit vector in the direction of $\mathbf{r}$, $K\left( a,b,z \right)$ is the Kummer function and the function $B$ is defined as $B\left(a,z\right)=-ia\operatorname{Im}\left(K\left(a+1,2,z\right)K\left(\bar{a},1,\bar{z} \right)\right)$ (the bar denotes complex conjugate).

To obtain the density and current distribution of an ensemble of DM particles, one needs to add up the contributions from all DM particles with different velocities in the ensemble. Since we assumed that the DM particles ensemble is non-interacting, the particle velocity $\mathbf{v}$ obeys the Maxwell - Boltzmann distribution
\begin{equation}\label{Boltzmann}
f\left( \mathbf{v} \right){{d}^{3}}v=\frac{{{e}^{-\frac{{{\left( \mathbf{v}-{{\mathbf{v}}_{s}} \right)}^{2}}}{2{{v}_{r}}^{2}}}}{{d}^{3}}v}{{{\left( 2\pi  \right)}^{\frac{3}{2}}}{{v}_{r}}^{3}}\,,
\end{equation}
where $v_s$ is the mean velocity of the gravitating body relative to the DM ensemble and $v_r$ is the root-mean-square velocity of the DM ensemble. Note that for \eqref{Boltzmann} to hold, we have further assumed that the DM ensemble is isothermal \cite{bahcall1984comparisons,caldwell1981jar,PhysRevD.33.889,alvarado1987dynamical} and that the presence of the body's gravitational potential does not greatly disturb the structure of the 'free' velocity distribution (roughly speaking, this amounts to the condition that a DM particle's kinetic energy $T=\mu v^2/2$ is significantly greater than its potential energy $V = GM\mu/r$; this condition holds for a large portion of the DM ensemble). In the isothermal model, $v_r \approx \SI[per-mode=symbol]{270}{\kilo\meter\per\second} = \SI{9e-4}{}$. For simplicity, we take $v_s=v_{\odot}\approx\SI{8e-4}{}$ (as mentioned above, this value is in the correct order of magnitude for the bodies in question).

The total particle density and current distribution of the DM ensemble are then given by
\begin{equation} \label{total density}
\bar{\rho}\left(\mathbf{r}\right) = \int{\rho\left(\mathbf{r},\mathbf{v}\right)f\left(\mathbf{v}\right)d^3v} \,,
\end{equation}
and
\begin{equation} \label{total current}
\bar{\mathbf{j}}\left(\mathbf{r}\right) = \int{\mathbf{j}\left(\mathbf{r},\mathbf{v}\right)f\left(\mathbf{v}\right)d^3v}\,.
\end{equation}

\subsection{DM density}
We first note that as $r\to \infty $, $\rho $ is essentially unity everywhere so $\bar{\rho }\left( \infty  \right)=1$. This corresponds to the normalization of the density in the void between astronomical objects to unity.

\begin{figure}
\centering
\includegraphics[width=0.3\textwidth]{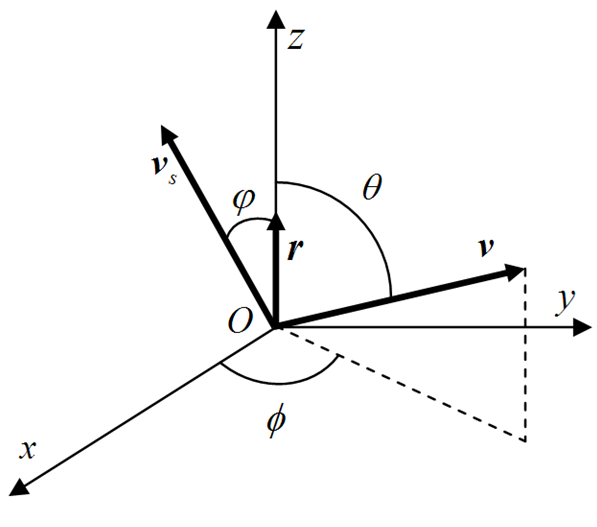}
\caption{Coordinates system used to evaluate the integrals \eqref{total density} and \eqref{total current}. Here, $0\leq\phi\leq 2\pi$ and $0\leq \theta \leq \pi$ are intergation variables whereas $\varphi$ is a free variable which determines the direction of observation $\mathbf{r}$.}\label{coordinates system}
\end{figure}

For finite $r$, let us set up the coordinates system as shown in Figure \ref{coordinates system}. The vector $\mathbf{r}$ points along the $z$-axis. The vector ${{\mathbf{v}}_{s}}$ lies in the $xz$-plane and forms an angle $\varphi $ with $\mathbf{r}$. We want to find $\bar{\rho }$ as a function of $r$ and $\varphi $ (clearly, by symmetry, these two are the only relevant parameters). 
With this setting, we have
%
%\begin{widetext}
\begin{equation} \label{rho integral with coordinates}
\begin{aligned}
  \bar{\rho }&=\Lambda\int {\left| K\left( \frac{i\kappa }{v},1,i\mu vr\left( 1-\cos \theta  \right) \right) \right|}^{2}  \\ 
  &\times \frac{{{e}^{\frac{-v^2+2v{{v}_{s}}\left( \sin \theta \cos \phi \sin \varphi +\cos \theta \cos \varphi  \right)}{2{{v}_{r}}^{2}}}}v}{{ 1-{{e}^{-\frac{2\pi \kappa }{v}}}}}  dv\, d\Omega \,,  \\ 
\end{aligned}
\end{equation}
%\end{widetext}
%
where $\Lambda =\kappa {{e}^{{{-v}_{s}}^{2}/{2{{v}_{r}}^{2}}}}/\sqrt{2\pi }{{v}_{r}}^{3}$, $d\Omega = \sin{\theta} \,d\theta\, d\phi$ and the integration limits are understood to be from $0$ to $\infty$ for $v$,  from $0$ to $\pi$ for $\theta$ and from $0$ to $2\pi$ for $\phi$.

It may be verified that the integrand of the integral \eqref{rho integral with coordinates} is less than or equal to ${{e}^{\frac{-{{v}^{2}}+2v{{v}_{s}}}{2{{v}_{r}}^{2}}}}v\left(1-{{e}^{-\frac{2\pi \kappa }{v}}}\right)^{-1}$, which, for all values of $\kappa$ below the bounds given in Table \ref{de Broglie mass}, is essentially zero for all $v$ outside the interval $\left[0.01v_s,10v_s\right]$. This observation allows us to cut off the $v$-integral in Eq.\ \eqref{rho integral with coordinates} at $v_{\text{min}}=0.01v_s$ and $v_{\text{max}}=10v_s$.
\subsection{DM current}
We first note that as $r \rightarrow \infty$, $\mathbf{j}\rightarrow\mathbf{v}$ so $\bar{\mathbf{j}}\rightarrow\mathbf{v}_s$.

For finite $r$, with a coordinates system as in Fig.\ \ref{coordinates system}, the components of the collective DM current $\bar{\mathbf{j}}$ can be written as
\begin{equation}\label{jx}
\begin{aligned}
 \bar{j}_x&=\Lambda \int \left[\left| K\left( \frac{i\kappa }{v},1,i\mu vr\left( 1-\cos \theta  \right) \right) \right|^{2}\right.\\
  &+ \left.B\left(\frac{i\kappa }{v},1,i\mu vr\left( 1-\cos \theta  \right)\right)\right]{\sin }\theta \cos\phi \\
  &\times \frac{{e}^{\frac{-{{v}^{2}}+2v{{v}_{s}}\left( \sin \theta \cos \phi \sin \varphi +\cos \theta \cos \varphi  \right)}{2{{v}_{r}}^{2}}}{v}^{2}}{1-{{e}^{-\frac{2\pi \kappa }{v}}}}
  dvd\Omega \,, \\ 
\end{aligned}
\end{equation}
and
\begin{equation}\label{jy}
\begin{aligned}
\bar{j}_y&=\Lambda \int \left[\left| K\left( \frac{i\kappa }{v},1,i\mu vr\left( 1-\cos \theta  \right) \right) \right|^{2}\right.\\
  &+ \left.B\left(\frac{i\kappa }{v},1,i\mu vr\left( 1-\cos \theta  \right)\right)\right]{\sin }\theta \sin\phi \\
  &\times \frac{{e}^{\frac{-{{v}^{2}}+2v{{v}_{s}}\left( \sin \theta \cos \phi \sin \varphi +\cos \theta \cos \varphi  \right)}{2{{v}_{r}}^{2}}}{v}^{2}}{1-{{e}^{-\frac{2\pi \kappa }{v}}}}
  dvd\Omega =0\,,\\
  \end{aligned}
\end{equation}
and
\begin{equation}\label{jz}
\begin{aligned}
  \bar{j}_z&=\Lambda \int\left[\left| K\left( \frac{i\kappa }{v},1,i\mu vr\left( 1-\cos \theta  \right) \right) \right|^{2}\right.\\
  &+ \left.B\left(\frac{i\kappa }{v},1,i\mu vr\left( 1-\cos \theta  \right)\right)\right]{\cos }\theta \\
  &\times \frac{{e}^{\frac{-{{v}^{2}}+2v{{v}_{s}}\left( \sin \theta \cos \phi \sin \varphi +\cos \theta \cos \varphi  \right)}{2{{v}_{r}}^{2}}}{v}^{2}}{1-{{e}^{-\frac{2\pi \kappa }{v}}}}
  dvd\Omega  \\ 
  &-\Lambda \int B\left(\frac{i\kappa }{v},1,i\mu vr\left( 1-\cos \theta  \right)\right)\\
  &\times \frac{{e}^{\frac{-{{v}^{2}}+2v{{v}_{s}}\left( \sin \theta \cos \phi \sin \varphi +\cos \theta \cos \varphi  \right)}{2{{v}_{r}}^{2}}}{v}^{2}}{1-{{e}^{-\frac{2\pi \kappa }{v}}}}
  dvd\Omega \,.\\
\end{aligned}
\end{equation}

We observe that the $y$-component of the collective current $\bar{\mathbf{j}}$ vanishes. Thus, $\bar{\mathbf{j}}$ lies in the plane defined by $\mathbf{r}$ and $\mathbf{v}_s$.

For the same reason as above, the $v$-integrals in Eq.\ \eqref{jx} and\ \eqref{jz} cat be cut off at $v_{\text{min}}=0.01v_s$ and $v_{\text{max}}=10v_s$.
\section{Results}
\subsection{DM density}
\subsubsection{DM density near Earth}
\begin{figure}
\centering
\includegraphics[width=0.4\textwidth]{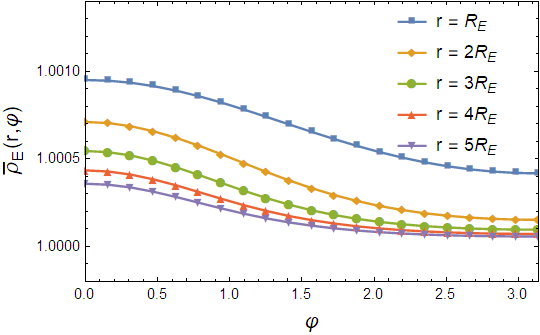}
\caption{The dependence of the DM density near Earth on the observation angle $\varphi$. The DM density here is the same as that at infinity.}\label{DensityEarth}
\end{figure}

Since the gravitational field created by Earth (mass $M_E\approx\SI{6e24}{\kilo\gram}$ and radius $R_E=\SI{6371}{\kilo\meter}$) is weak compared to most astronomical bodies, one expects that the DM ensemble will not be greatly disturbed. As a result, the \textit{single} particle density \eqref{one particle density} should be essentially unity everywhere and thus the \textit{collective} DM density is
\begin{equation}
\bar{\rho}\left(\mathbf{r}\right) \approx \int{f\left(\mathbf{v}\right)d^3v} =1
\end{equation}
This heuristic result agrees with direct numerical integration of Eq.\ \eqref{rho integral with coordinates}. Here and below, all numerical calculations are performed with some simple \textit{Mathematica} codes. The dependence of the DM density at the surface of Earth on $\varphi$ for some values of $r$, as calculated with such code, is shown in Fig.\ \ref{DensityEarth}. 

\subsubsection{DM density near Sun}
\begin{figure}
\centering
\includegraphics[width=0.4\textwidth]{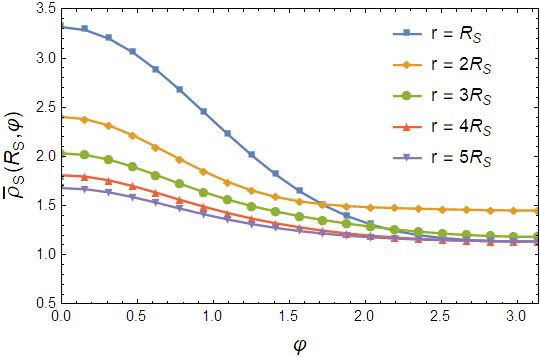}
\caption{The dependence of the DM density near Sun on the observation angle $\varphi$ for some values of the observation distance $r$. At Sun's surface there is an enhancement by a factor of 1.5 to 3.5 compared to the DM density at the surface of Earth.}\label{DensitySun}
\end{figure}

The dependence of the DM density near Sun (mass $M_S \approx \SI{2e30}{\kilo\gram}$ and radius $R_S\approx\SI{7e5}{\kilo\meter}$) on $\varphi$ for some values of $r$ is shown in Fig.\ \ref{DensitySun}. Evidently, at the surface of Sun, the DM density is from $1.5$ to $3$ times larger than that near Earth.

\subsubsection{DM density near a Neutron Star}
In the case of a typical Neutron Star (mass $M_{NS} \approx 2M_S$ and radius $R_{NS} \approx \SI{10}{\kilo\meter}$), direct numerical integration of Eq.\ \eqref{rho integral with coordinates} converges very slowly. This calls for certain approximations to be made. In Appendix \ref{AppendixA}, we demonstrate that if the condition
\begin{equation}
\nonumber
\mu r < \frac{\sqrt{100v_s^2+4\kappa^2}}{50v_s^2}
\end{equation}
is met, such as in the case of a typical Neutron Star or a White Dwarf, one can reduce Eq.\ \eqref{rho integral with coordinates} to a more integrable form
\begin{equation}\label{neutron star integral}
\begin{aligned}
\bar{\rho }&\approx2\Lambda \sqrt{\frac{2}{\kappa \mu r}} \int\limits_{v_{\rm min}}^{v_{\rm max}}\int\limits_{0}^{\frac{\pi }{2}} {{{e}^{\frac{-{{v}^{2}}+2v{{v}_{s}}\cos \varphi \cos 2x}{2{{v}_{r}}^{2}}}}}\\
&\times {{I}_{0}}\left( \frac{v{{v}_{s}}\sin \varphi \sin 2x}{{{v}_{r}}^{2}} \right)v\cos xdvdx \,,
\end{aligned}
\end{equation} 
where $I_0\left(x\right)$ is the zeroth-order modified Bessel function of the first kind.

\begin{figure}
\centering
\includegraphics[width=0.4\textwidth]{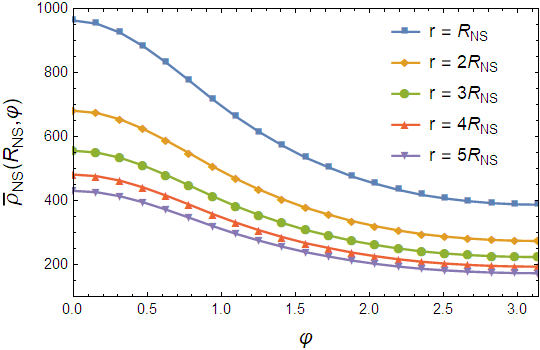}
\caption{The dependence of the DM density near a typical Neutron Star on the observation angle $\varphi$. At the Neutron Star's surface, there is an enhancement of two to three orders of magnitude compared to the DM density near Earth.}\label{DensityNeutron}
\end{figure}

The results of integrating of Eq.\ \eqref{neutron star integral} for some values of $r$ are shown in Fig.\ \ref{DensityNeutron}. We observe that near a typical Neutron Star, the DM density is enhanced by two to three orders of magnitude compared to that near Earth.

\subsubsection{DM density near a White Dwarf}
\begin{figure}
\centering
\includegraphics[width=0.4\textwidth]{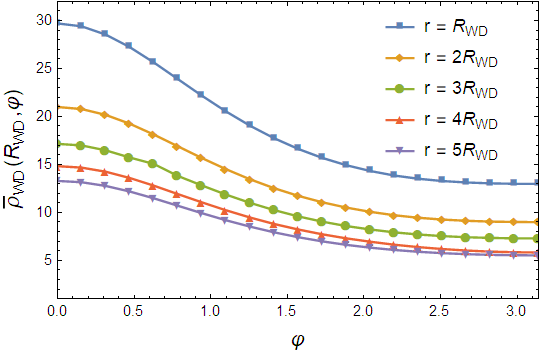}
\caption{The dependence of the DM density near typical White Dwarf on the observation angle $\varphi$. At the White Dwarf's surface, there is an enhancement by a factor of $15$ to $30$ compared to the DM density at the surface of Earth.}\label{DensityWhiteDwarf}
\end{figure}

In the case of a typical White Dwarf (mass $M_{WD} \approx M_S$ and radius $R_{WD} \approx R_E$), direct integration of Eq.\ \eqref{rho integral with coordinates} converges reasonably fast and the results for some values of $r$ are shown in Fig.\ \ref{DensityWhiteDwarf}.

\begin{figure}
\centering
\includegraphics[width=0.4\textwidth]{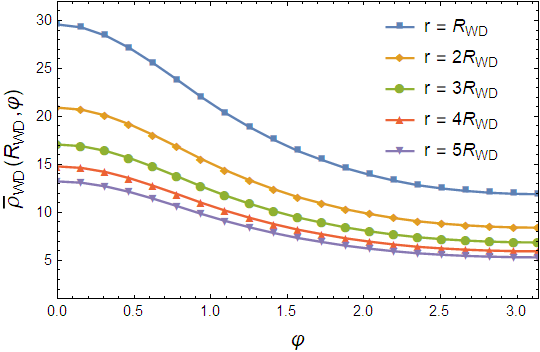}
\caption{The (approximate) dependence of DM density at the surface of a typical White Dwarf on the observation angle $\varphi$.}\label{DensityApproxWhiteDwarf}
\end{figure}

As discussed in Appendix \ref{AppendixA}, the approximations used in the calculation for a typical Neutron Star also apply to that for a typical White Dwarf. These approximations simplifies Eq.\ \eqref{rho integral with coordinates} to Eq.\ \eqref{neutron star integral}, which when integrated converges much faster than the former. The results of integrating Eq.\ \eqref{neutron star integral} (in the case of a White Dwarf) for some values of $r$ are shown in Fig.\ \ref{DensityApproxWhiteDwarf}. These agree, within a few percents, with those obtained by integrating the exact Eq.\ \eqref{rho integral with coordinates}.
\subsubsection{Comparison with the classical results}
\begin{figure}
\centering
\includegraphics[width=0.4\textwidth]{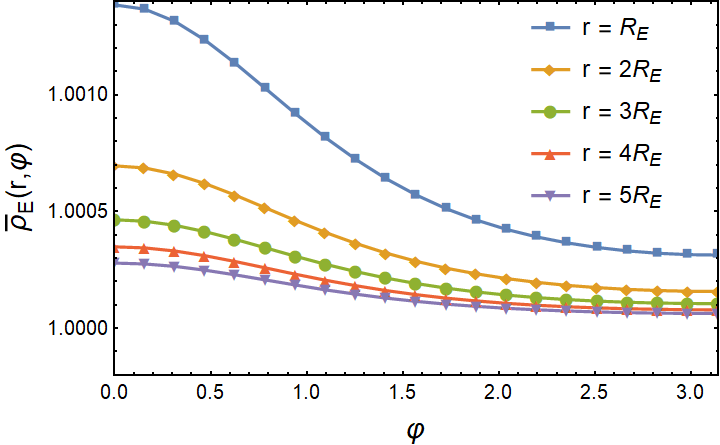}
\caption{The dependence of the \textit{classical} DM density near Earth on the observation angle $\varphi$.}\label{EClassical}
\end{figure}

\begin{figure}
\centering
\includegraphics[width=0.4\textwidth]{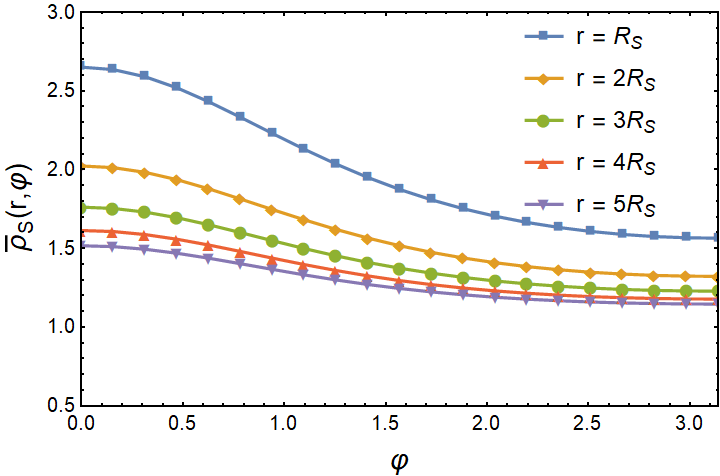}
\caption{The dependence of the \textit{classical} DM density near Sun on the observation angle $\varphi$.}\label{SClassical}
\end{figure}

\begin{figure}
\centering
\includegraphics[width=0.4\textwidth]{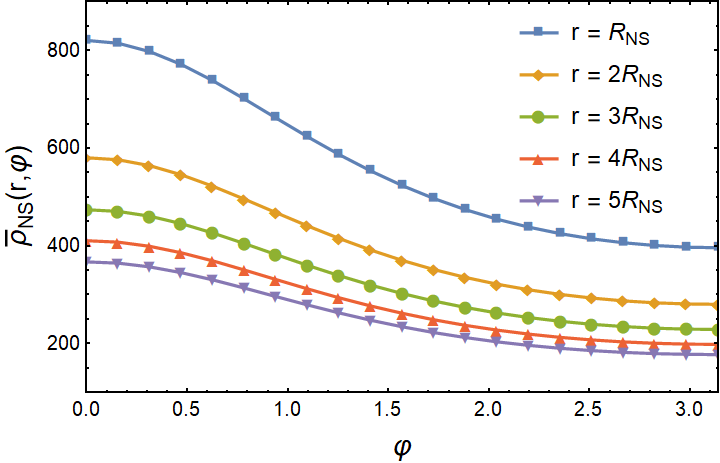}
\caption{The dependence of the \textit{classical} DM density near a typical Neutron Star on the observation angle $\varphi$.}\label{NSClassical}
\end{figure}

\begin{figure}
\centering
\includegraphics[width=0.4\textwidth]{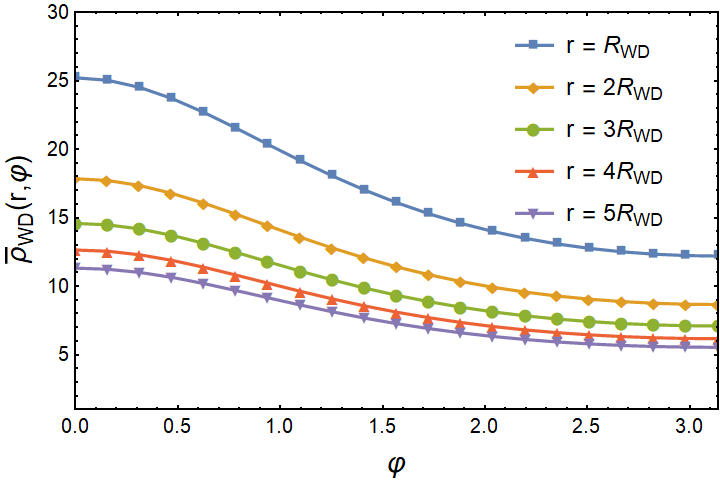}
\caption{The dependence of the \textit{classical} DM density near a typical White Dwarf on the observation angle $\varphi$.}\label{WDClassical}
\end{figure}

As mentioned before, the distribution of a DM ensemble of mass $\gtrsim \mu\SI{}{\electronvolt}$ can be studied in the framework of classical mechanics. The case of such ensemble near Sun was examined in \cite{SikivieWick2002}. The extension of this examination to the case of Earth, Neutron Stars and White Dwarfs can be readily carried out. The results are plotted in Figs.\ \ref{EClassical},\ \ref{SClassical},\ \ref{NSClassical} and\ \ref{WDClassical}. Comparing these with Figs.\ \ref{DensityEarth},\ \ref{DensitySun},\ \ref{DensityNeutron} and\ \ref{DensityWhiteDwarf}, one observes that the classical and quantum results agree, at least in the order of magnitude and behavior of the curves (the disagreement in the actual numerical factors is to be expected due to the change of DM particle's nature in the gravitational potential from classical to quantum).
\subsection{DM current}
\subsubsection{DM current near Earth}
\begin{figure}
\centering
\includegraphics[width=0.4\textwidth]{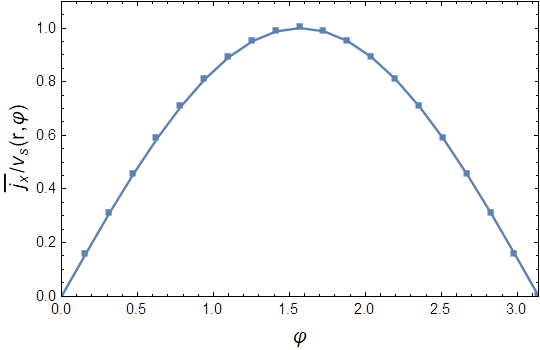}
\caption{The dependence of the $x$-component of the DM current near Earth on the observation angle $\varphi$. This dependence is the same for all value of $r$.}\label{jxE}
\end{figure}

\begin{figure}
\centering
\includegraphics[width=0.4\textwidth]{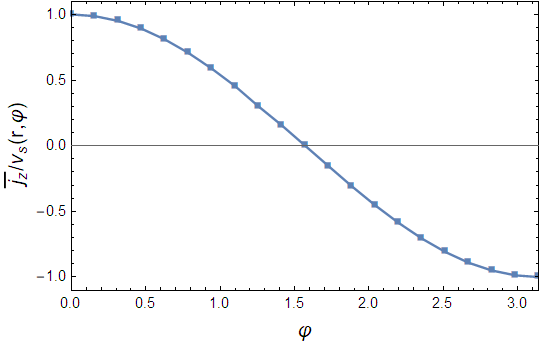}
\caption{The dependence of the $z$-component of the DM current near Earth on the observation angle $\varphi$. This dependence is the same for all value of $r$. It is evident that near Earth, $\bar{j}_x=v_s\cos\varphi=\left(v_s\right)_x$}\label{jzE}
\end{figure}

Since the gravitational field of Earth is weak, we expect the DM current near Earth to be the same as that in the void between astronomical objects, that is $\bar{\mathbf{j}}_{r \geq R_E} \approx \mathbf{v}_s$. This heuristic result is confirmed by direct integration of Eqs.\ \eqref{jx} and\ \eqref{jz}. The dependence of $\bar{j}_x/v_s$ and $\bar{j}_z/v_s$ near Earth on $\varphi$ is shown in Figs.\ \ref{jxE} and\ \ref{jzE} (these plots are the same for all values of $r$). Clearly, $j_x/v_s=\sin\varphi$ and $j_x/v_s=\cos\varphi$, as expected.

\subsubsection{DM current near Sun}
\begin{figure}
\centering
\includegraphics[width=0.4\textwidth]{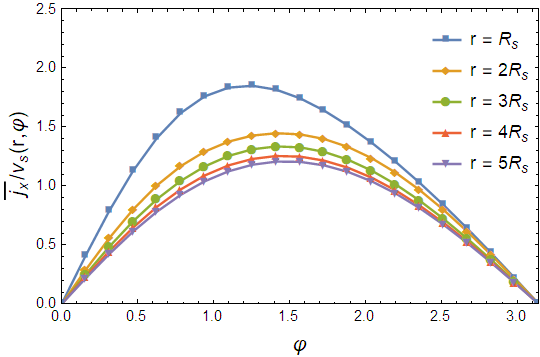}
\caption{The dependence of the $x$-component of the DM current near Sun on the observation angle $\varphi$ for some values of $r$.}\label{jxS}
\end{figure}

\begin{figure}
\centering
\includegraphics[width=0.4\textwidth]{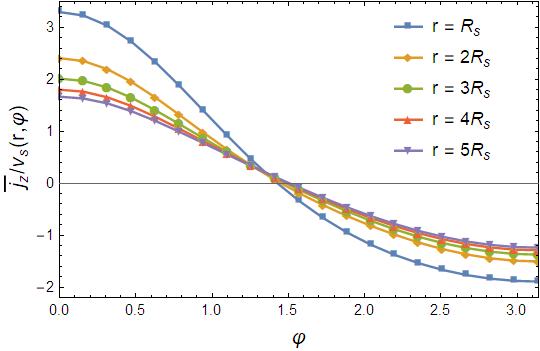}
\caption{The dependence of the $z$-component of the DM current near Sun on the observation angle $\varphi$ for some values of $r$.}\label{jzS}
\end{figure}

The dependence of $\bar{j}_x/v_s$ and $\bar{j}_z/v_s$ near Sun on $\varphi$ for some values of $r$ is shown in Figs.\ \ref{jxS} and\ \ref{jzS}. Evidently, at the surface of Sun, the DM current is from 2 to 3 times stronger than that near Earth. One also observes some deviations of the direction of the current from that of $\mathbf{v}_s$.

\subsubsection{DM current near a Neutron Star}
In the case of a typical Neutron Star, direct integration of Eqs.\ \eqref{jx} and\ \eqref{jz} converges very slowly. Instead, just like before, one can use the approximations discussed in the Appendix to simplify Eqs.\ \ref{jx} and\ \ref{jz} to more integrable forms
\begin{equation}\label{jx approx}
\begin{aligned}
\bar{j}_x&\approx\frac{\Lambda }{\pi }\sqrt{\frac{2}{\kappa \mu r}}\int\limits_{0}^{2\pi }\int\limits_{v_{\rm min}}^{{v}_{\rm max}}\int\limits_{0}^{\frac{\pi }{2}}{{{v}^{2}}\cos x \sin 2x \cos\phi} \\
&\times {e}^{\frac{-{{v}^{2}}+2v{{v}_{s}}\left( \cos \varphi \cos 2x+\cos \phi \sin \varphi \sin 2x \right)}{2{{v}_{r}}^{2}}}d\phi dvdx \,,
\end{aligned}
\end{equation}
and
\begin{equation}\label{jz approx}
\begin{aligned}
\bar{j}_z &\approx 2\Lambda \sqrt{\frac{2}{\kappa \mu r}}\int\limits_{v_{\rm min}}^{v_{\rm max}}\int\limits_{0}^{\frac{\pi }{2}}{I}_{0}\left( \frac{v{{v}_{s}}\sin \varphi \sin 2x}{{{v}_{r}}^{2}} \right) \\
&\times {e}^{\frac{-{{v}^{2}}+2v{{v}_{s}}\cos \varphi \cos 2x}{2{{v}_{r}}^{2}}}{{v}^{2}}\cos x\cos 2xdvdx \,.
\end{aligned}
\end{equation}

\begin{figure}
\centering
\includegraphics[width=0.4\textwidth]{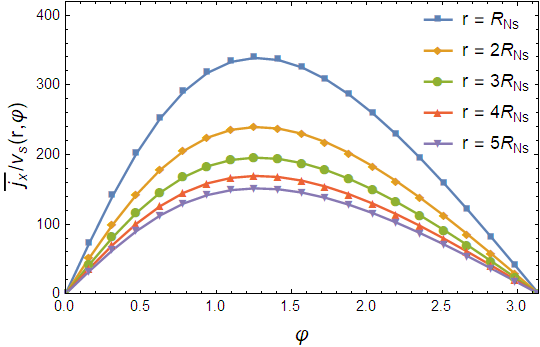}
\caption{The dependence of the $x$-component of the DM current near a typical Neutron Star on the observation angle $\varphi$ for some values of $r$.}\label{jxNS}
\end{figure}

\begin{figure}
\centering
\includegraphics[width=0.4\textwidth]{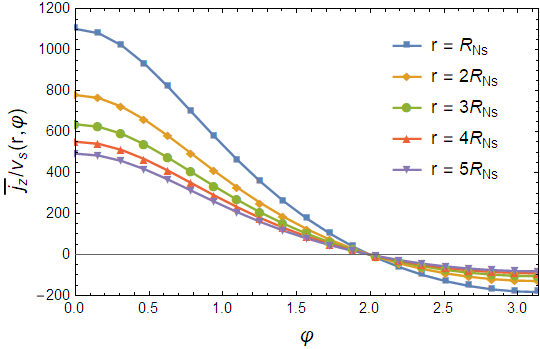}
\caption{The dependence of the $z$-component of the DM current near a typical Neutron Star on the observation angle $\varphi$ for some values of $r$.}\label{jzNS}
\end{figure}

The results of integrating these two equations for the case of a typical Neutron Star are shown in Figs.\ \eqref{jxNS} and\ \eqref{jzNS}. We observe that at the surface of the Neutron Star, the DM current is enhanced by two to three orders of magnitude compared to the current near Earth.

\subsubsection{DM current near a White Dwarf}
\begin{figure}
\centering
\includegraphics[width=0.4\textwidth]{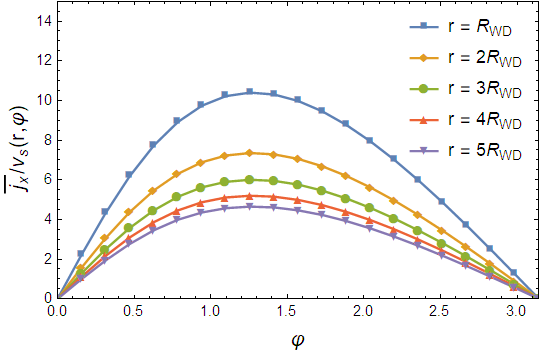}
\caption{The dependence of the $x$-component of the DM current near a typical White Dwarf on the observation angle $\varphi$ for some values of $r$.}\label{jxWD}
\end{figure}

\begin{figure}
\centering
\includegraphics[width=0.4\textwidth]{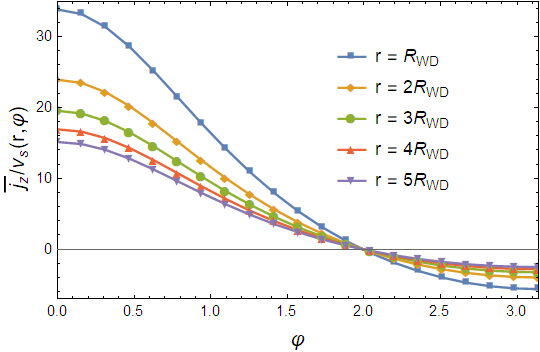}
\caption{The dependence of the $z$-component of the DM current near a typical White Dwarf on the observation angle $\varphi$ for some values of $r$.}\label{jzWD}
\end{figure}

Direct integration of the DM current near a typical White Dwarf converges very slowly. One must, therefore, resort to the simplified formulae\ \eqref{jx approx} and\ \eqref{jz approx} for computation. The results of integrating these equations in the case of a White Dwarf are shown in Figs.\ \eqref{jxWD} and\ \eqref{jzWD}. 

\section{Discussion} \label{Discussion}
We have shown that near some compact astronomical objects, there are significant enhancements of lightweight DM density and current compared to those near Earth. Specifically, near Sun, the density and current are about two to three times larger than those near Earth; near a typical White Dwarf, they are about ten to thirty times larger and near a Neutron star, the enhancements are from two to three orders of magnitudes.

It is worth noting that the magnitudes of these enhancements are insensitive to the DM mass (as to be expected, at least in classical mechanics), as long as it does not exceeds the bounds presented in Table\ \ref{de Broglie mass}. This fact can be verified numerically for the case of Earth and Sun. For other cases, it can be verified directly from Eqs.\ \eqref{rho_approx},\ \eqref{jx approx} and\ \eqref{jz approx}. 

Finally, we provide an example where the results of this paper may prove useful. The positions of the lines in the atomic absorption and emission spectra observed from any astronomical onject depend on the value of the fine structure constant $\alpha$ \cite{griffiths2005introduction}, which, as conjectured in \cite{stadnik2015can}, can change due to the DM-SMM interactions. This change in $\alpha$ may depend on the DM density \cite{stadnik2015can}. It is therefore possible to detect DM-induced variations of $\alpha$ by analyzing spectra of Neutron Stars of White Dwarfs.

Neutron star spectrum data are available from observations by Cottam \textit{et al} \cite{cottam2002gravitationally,cottam2008burst}. These data appear to be very suitable for our calculation since they imply a gravitational redshift of $z=0.35$, which is consistent with most modern equations of state for Neutron Stars in the mass range of $1.4-1.8 M_{\odot}$ and $R \sim 9-12 \si{km}$. Unfortunately, if one assumed that the shifts of lines (after the effect of gravitational redshift has been eliminated by taking the ratios of the line wavelengths) are due solely to variation of $\alpha$, one obtains an unrealistic large value of $\alpha$ ($\approx 1/16$). One possible reason for this is that near a Neutron Star, there exist large magnetic fields which can induce significant spectral lines shifts. One should therefore consider some situation where magnetic fields are absent, e.g., in the vicinity of stars that are closest to the supermassive black hole Sagittarius A* at the centre of the Milky Way. The situation there is slightly different from that considered above: the DM concentration near these stars is caused mainly by the black hole's gravitational field (since the black hole is more massive than the stars). Hence, although the mass that enters our calculations is that of the black hole, the radii should be the distances from the black hole to the stars orbiting it. Applying the calculations above the cases of the stars S1 and S14, which can come very close to the black hole, with pericentral distances 121 au and 109 au, respectively, one obtains enhancements of the DM density and current of about 15 to 40 times. It is thus possible to detect DM-induced variation of $\alpha$ by analyzing spectra coming from these stars. To the authors' knowledge, no such spectra are currently available.
\section*{Acknowledgments}
The authors thank Igor Samsonov of helpful discussions.

\section*{Appendix: Approximate formulae for DM density and current near a compact object}\label{AppendixA}

In this Appendix, we prove that for certain compact (defined below) objects, one is allowed to make certain approximations which reduce Eqs.\ \eqref{rho integral with coordinates},\ \eqref{jx} and\ \eqref{jz} to the simpler Eqs.\ \eqref{neutron star integral},\ \eqref{jx approx} and\ \eqref{jz approx} which have the advantage of having a fast-converging numerical integration. 

The approximations employed involve the large parameter asymptotic form of the Kummer function $K\left(a,b,z\right)$ with $b\geq 1$, to which we give a brief summary here. Recall that the Kummer function $K\left(a,b,z\right)$ is a solution to the differential equation
\begin{equation} 
z\frac{d^2K}{dz^2}+\left(b-z\right)\frac{dK}{dz}- a K=0\,,
\end{equation}
with some boundary conditions.

By setting $v=e^{-z/2}\left[\left(b/2-a\right)z\right]^{b/2-1/2}K$ and $x=\sqrt{2z\left(b-2a\right)}$ one obtains the following differential equation of $v$ with respect to $x$
\begin{equation}\label{v differential eqn}
\frac{d^2 v}{dx^2}+\frac{1}{x}\frac{dv}{dx}+\left[1-\frac{\left(b-1\right)^2}{x^2}-\frac{z}{2\left(b-2a\right)}\right]v=0\,.
\end{equation}

If $\left|z\right| \ll \left|2\left(b-2a\right)\right|$ then Eq.\ \eqref{v differential eqn} becomes the Bessel differential equation with solution $v=J_b\left(x\right)$. Thus, the function $K\left(a,b,z\right)$ has the asymptotic form
\begin{equation}\label{asymptotic form}
\begin{aligned}
K\left(a,b,z\right)\xrightarrow{\left|z\right| \ll \left|2\left(b-2a\right)\right|}\\
\Gamma\left(b\right)e^{\frac{z}{2}}\left[\left(b/2-a\right)z\right]^{\frac{1-b}{2}}J_b\left(\sqrt{2z\left(b-2a\right)}\right)\,,
\end{aligned}
\end{equation}
where coefficient $\Gamma\left(b\right)$ is present to guarantee the correct normalization of the Kummer function.

In the case of the function $K\left( \frac{i\kappa }{v},1,i\mu vr\left( 1-\cos \theta  \right) \right)$, the condition for the applicability of Eq.\ \eqref{asymptotic form} reads
\begin{equation}
\mu r \ll \frac{2\sqrt{v^2+4\kappa^2}}{v^2}\,,
\end{equation}
which holds for all $0.01v_s \leq v \leq 10v_s$ (recall that we cut off the $v$-integral in Eq.\ \eqref{rho integral with coordinates} at these limits) iff
\begin{equation}\label{applicability condition}
\mu r < \frac{\sqrt{100v_s^2+4\kappa^2}}{50v_s^2}\,.
\end{equation}
Condition \eqref{applicability condition} is satisfied in the case of a typical Neutron Star for $r \lesssim 10^4R_{NS} \approx \SI{e5}{\kilo\meter}$ and in the case of a typical White Dwarf for $r \lesssim 11_{WD} \approx \SI{e5}{\kilo\meter}$. This condition is, however, not met in the case of Earth or Sun.

In the cases that condition \eqref{applicability condition} is satisfies, using the asymptotic form \eqref{asymptotic form} in Eq.\ \eqref{rho integral with coordinates} yields
\begin{equation}
\begin{aligned}
\bar{\rho }\approx\Lambda \int\limits_{0}^{2\pi }\int\limits_{0}^{\pi }\int\limits_{v_{\rm min}}^{v_{\rm max}}{{J}_{0}}{{\left( 2\sqrt{\kappa \mu r\left( 1-\cos \theta  \right)} \right)}^{2}} \\
\times \frac{{{e}^{\frac{-v^2+2v{{v}_{s}}\left( \sin \theta \cos \phi \sin \varphi +\cos \theta \cos \varphi  \right)}{2{{v}_{r}}^{2}}}}v}{{ 1-{{e}^{-\frac{2\pi \kappa }{v}}}}}  dv\sin{\theta} \,d\theta\, d\phi  \,,
\end{aligned}
\end{equation}
and a change of variable $u=1-\cos \theta $ turns this into
%
%\begin{widetext}
\begin{equation}\label{rho_approx}
\begin{gathered}
\bar{\rho }\approx\Lambda \int\limits_{0}^{2\pi }{d\phi \int\limits_{v_{\rm min}}^{v_{\rm max}}{{{e}^{\frac{-{{v}^{2}}+2v{{v}_{s}}\cos \varphi }{2{{v}_{r}}^{2}}}}vdv}} \\
\times \int\limits_{0}^{2}{{{J}_{0}}{{\left( 2\sqrt{\kappa \mu ru} \right)}^{2}}{{e}^{\frac{v{{v}_{s}}\left[ \sqrt{2u-{{u}^{2}}}\cos \phi \sin \varphi -u\cos \varphi \right]}{{{v}_{r}}^{2}}}}du}\,.
\end{gathered}
\end{equation}
%\end{widetext}
%
Note that we have discarded the term $e^{-\frac{2\pi\kappa}{v}}$ in the denominator since it is very small in the case of a Neutron Star or a White Dwarf.

The integral $I=\int\limits_{0}^{2}{{{J}_{0}}{{\left( \beta \sqrt{u} \right)}^{2}}{{e}^{-\alpha u+\chi \sqrt{2u-{{u}^{2}}}}}du} $ where $\alpha =\frac{v{{v}_{s}}\cos \varphi }{{{v}_{r}}^{2}}$, $\beta =2\sqrt{\kappa \mu r}$ and $\chi =\frac{v{{v}_{s}}\cos \phi \sin \varphi }{{{v}_{r}}^{2}}$ can be estimated as follows. 

Since the $v$-integral is cut off at $10v_s$, $\alpha$ and $\chi$ is at most of the order of $10$. For a typical Neutron Star or a typical White Dward, $\beta \gg 10$. As a result, one can write
%
%\begin{widetext}
\begin{equation}\
\begin{aligned}
	I=\underbrace{\int\limits_{0}^{\delta /{{\beta }^{2}}}{{{J}_{0}}{{\left( \beta 				\sqrt{u} \right)}^{2}}{{e}^{-\alpha u+\chi \sqrt{2u-{{u}^{2}}}}}du}}_{{{I}_{1}}}\\
	+\underbrace{\int\limits_{\delta /{{\beta }^{2}}}^{2}{{{J}_{0}}{{\left( \beta 				\sqrt{u} \right)}^{2}}{{e}^{-\alpha u+\chi \sqrt{2u-{{u}^{2}}}}}du}}_{{{I}_{2}}}\,,
	\end{aligned}
\end{equation}
%\end{widetext}
%
where $0\leq \delta \leq 2\beta^2$ is a real number chosen such that one can, in ${{I}_{1}}$, use the small argument expansion of ${{e}^{-\alpha u+\chi \sqrt{2u-{{u}^{2}}}}}$ and, in ${{I}_{2}}$, use the large argument asymptotic form of ${{J}_{0}}{{\left( \beta \sqrt{u} \right)}^{2}}$. To the lowest order in $\beta^{-1}$, one thus has
%
%\begin{widetext}
\begin{equation}
	{{I}_{1}}\approx 0 \,,
\end{equation}
%\end{widetext}
%
and
\begin{equation}
  {{I}_{2}}\approx \frac{2}{\beta }\int\limits_{\delta/{\beta }^{2}}^{2}{\frac{{{e}^{-\alpha u+\chi \sqrt{2u-{{u}^{2}}}}}{{\cos }^{2}}\left( \beta \sqrt{u}-\frac{\pi }{4} \right)}{\pi \sqrt{u}}du}\,.
\end{equation}
%\end{widetext}
%
Since the cosine function oscillates rapidly, one can replace it with $\frac{1}{2}$ and obtain
\begin{equation}
I\approx \frac{1}{\pi \beta }\int\limits_{\delta/{\beta }^{2}}^{2}{\frac{{{e}^{-\alpha u+\chi \sqrt{2u-{{u}^{2}}}}}}{\sqrt{u}}du}\,,
\end{equation}
which, with a change of variable  $u=1-\cos 2x$ gives
\begin{equation}
  I\approx \frac{2\sqrt{2}}{\pi \beta }\int\limits_{0}^{\frac{\pi }{2}}{{{e}^{-2\alpha {{\sin }^{2}}x+\chi \sin 2x}}\cos xdx}\,,
\end{equation}
so Eq.\ \eqref{rho_approx} becomes
\begin{equation}
\begin{aligned}
\bar{\rho }\approx2\Lambda \sqrt{\frac{2}{\kappa \mu r}} \int\limits_{v_{\rm min}}^{v_{\rm max}}\int\limits_{0}^{\frac{\pi }{2}} {{{e}^{\frac{-{{v}^{2}}+2v{{v}_{s}}\cos \varphi \cos 2x}{2{{v}_{r}}^{2}}}}}\\
\times {{I}_{0}}\left( \frac{v{{v}_{s}}\sin \varphi \sin 2x}{{{v}_{r}}^{2}} \right)v\cos xdxdv \,,
\end{aligned}
\end{equation}
which is the needed Eq.\ \eqref{neutron star integral}.

In deriving the approximate formulae\ \eqref{jx approx} and\ \eqref{jz approx}, we note that by using the large parameter asymptotic form \eqref{asymptotic form} of the Kummer functions, one obtains
\begin{equation}
\begin{aligned}
  K\left( \frac{i\kappa }{v}+1,2,i\mu vr\left( 1-\cos \theta  \right) \right)\\
  \times K\left( -\frac{i\kappa }{v},1,-i\mu vr\left( 1-\cos \theta  \right) \right)\\
  \approx \frac{{{J}_{1}}\left( 2\lambda  \right){{J}_{0}}\left( 2\lambda  \right)}{\lambda } 
  \end{aligned}\,,
\end{equation}
where $\lambda =\sqrt{\kappa \mu r\left( 1-\cos \theta  \right)}$.
Clearly this is real so we have
%
%\begin{widetext}
\begin{equation}
B\left(\frac{i\kappa}{v},i\mu vr\left(1-\cos\theta\right)\right)=0\,,
\end{equation}
so Eq.\ \eqref{one particle current} reduces to
\begin{equation}
\mathbf{j} \approx \rho\mathbf{v}\,,
\end{equation}
From this, formulae\ \eqref{jx approx} and\ \eqref{jz approx} follow readily.

\bibliography{Ref}

\end{document}